\documentclass[a4paper]{jpconf}

\usepackage{graphicx}
\usepackage{epsfig}
\usepackage{multicol}
\usepackage{pstricks,pst-grad}
\usepackage{multirow}
\usepackage{subfigure}

\begin{document}
\title{State-space modelling for heater induced thermal effects on LISA Pathfinder's Test Masses}

\author{F Gibert$^{1,2}$, M Nofrarias$^{1,2}$, M Diaz-Aguil\'o$^{1,3}$, A Lobo$^{1,2}$, N Karnesis$^{1,2}$, I Mateos$^{1,2}$, J Sanju\'an$^4$, I Lloro$^{1,2}$, L Gesa$^{1,2}$ and V Mart\'in$^{1,2}$ }
\address{$^1$ Institut d'Estudis Espacials de Catalunya (IEEC), Barcelona, Spain}
\address{$^2$ Institut de Ci\`encies de l'Espai (CSIC-IEEC), Barcelona, Spain}
\address{$^3$ Escola Polit\`ecnica Superior de Castelldefels (EPSC), Universitat Polit\`ecnica de Catalunya (UPC), Castelldefels, Spain}
\address{$^4$ Department of Physics, University of Florida, Gainesville, FL, United States}

\ead{gibert@ieec.cat}

\begin{abstract}
The OSE (Offline Simulations Environment) simulator of the LPF (LISA Pathfinder) mission is intended to simulate the different experiments to be carried out in flight. Amongst these, the thermal diagnostics experiments are intended to relate thermal disturbances and interferometer readouts, thereby allowing the subtraction of thermally induced interferences from the interferometer channels. In this paper we report on the modelling of these simulated experiments, including the parametrisation of different thermal effects (radiation pressure effect, radiometer effect) that will appear in the Inertial Sensor environment of the LTP (LISA Technology Package). We report as well how these experiments are going to be implemented in the LTPDA toolbox, which is a dedicated tool for LPF data analysis that will allow full traceability and reproducibility of the analysis thanks to complete recording of the processes.
\end{abstract}

\section{Introduction}
\label{intro}
The Thermal Diagnostic Subsystem of the LISA Technology Package (LTP) onboard LISA Pathfinder (LPF)~\cite{lpf1, lpf2} includes the generation of thermal disturbances in some thermal sensitive locations of the LTP to induce perturbing signals in the interferometer (IFO) readouts. The objective of such controlled perturbations is to obtain information of the effects that potential temperature gradients in the LTP may induce in the final IFO readout during the main operational phase, enabling the possibility of removing thermally induced noise from the signal.\\

The LISA Pathfinder main noise requirements in the LTP state for the test masses residual differential acceleration:
  \begin{eqnarray}
  	\label{reqeq}
    S^{1/2}_{\Delta a,\, {\rm LPF}}(\omega)\leq 3\times10^{-14}\left[1+\left(\frac{\omega/2\pi}{3\, {\rm mHz}}\right)^{2}\right]\, {\rm m}\, {\rm s}^{-2}\, {\rm Hz}^{-1/2} 
  \end{eqnarray}
  in the bandwidth between 1\,mHz and 30\,mHz~\cite{reqsSpec} .\\

Therefore, a set of Thermal Diagnostics experiments consisting of the injection of controlled loads of heat at specific spots must be carefully designed, producing in-band perturbing signals with high Signal-to-Noise Ratio ($SNR \approx 50$). Moreover, a model linking both the heat propagation and temperature distributions to final IFO readouts must be designed, and it is intended to be introduced into two LPF key simulators:
\begin{itemize}
\item The OSE ({\it Offline Simulation Environment}) simulator: Located at ESAC ({\it European Space Astronomy Centre}), this simulator is intended to validate all LPF mission operations considering the different interfaces and data transfer processes.
\item The LTPDA simulator: Developed by the LTP Data Analysis (LTPDA) team, within the LTPDA software~\cite{ltpdaref}. It is a linear simulator based on State-Space Modelling (SSM) intended to analyse the data delivered by the mission.
\end{itemize}

Part of this model is presented here in this paper in Section~\ref{modelling}, and some first results are reported in Section~\ref{results}. Nevertheless, it is important to first introduce the different Thermal Diagnostics experiments and the architecture of the Data Analysis associated to them.

\subsection{Thermal Diagnostics Experiments}
\label{experiments}
\begin{figure}
  \begin{center}
    \includegraphics[width=13cm]{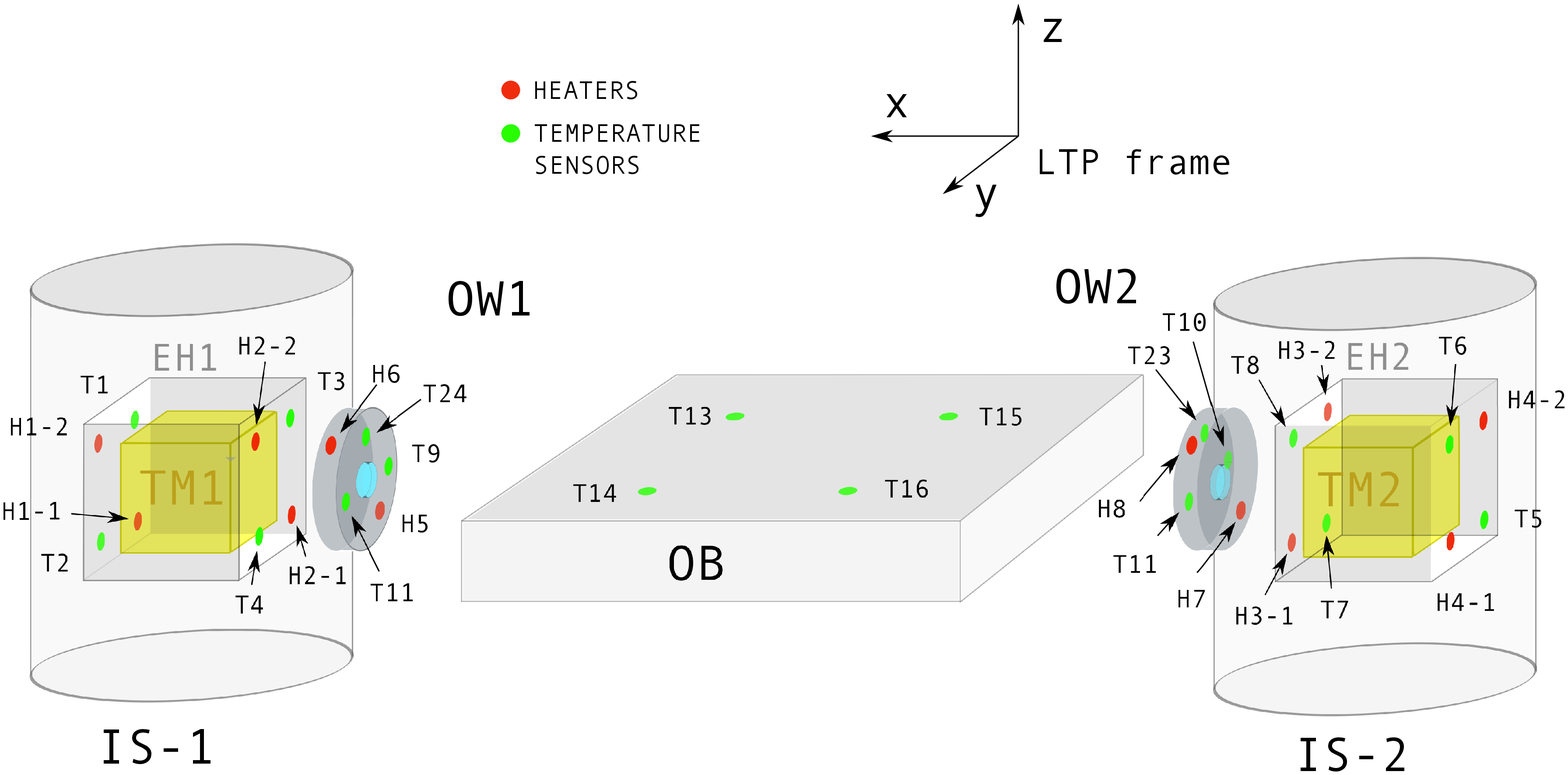}    
    \caption{\label{is_dds_items}Scheme of the Thermal DS (Diagnostics Subsystem) items locations in the Electrode Housings, on the Optical Windows and on the Optical Bench, with their enumeration.}
  \end{center}
\end{figure}

\begin{figure}
  \begin{center}
    \includegraphics[width=9cm]{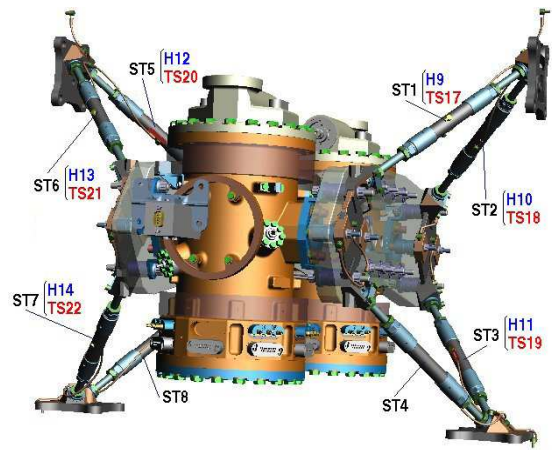}    
    \caption{\label{SSlayout}Scheme of the Thermal DS items location on the Suspension Struts. CAD rendering of the LTP courtesy of EADS Astrium GmbH.}
  \end{center}
\end{figure}

The LTP is equipped with a set of 14 heaters and 24 temperature sensors (see Figures~\ref{is_dds_items}~and~\ref{SSlayout}). Different experiments are intended to apply heat signals through the heaters and record the temperature readouts. These experiments are gathered in the Experiment Master Plan (EMP)~\cite{emp} which aims to characterise the thermal effects in Table~\ref{thermexp} by switching on different heaters at each case, and consequently thermally disturbing different parts of the LTP.

\begin{table}[h]
\caption{\label{thermexp}Types of Thermal Diagnostics Experiments. Heaters, associated thermal effects and expected main consequences.}
\begin{center}
\begin{tabular}{lll}
\br
Heaters activated & \hspace{1cm}Thermal effects & \hspace{1cm}Main consequence \\
\mr
\multirow{2}{*}{Electrode Housing (EH)} & \hspace{1cm}Radiometer effect, & \hspace{1cm}Force and torques  \\
& \hspace{1cm}radiation pressure  &  \hspace{1cm}on Test Masses (TMs)\\
& & \\
\multirow{2}{*}{Optical Window (OW)} & \hspace{1cm}Glass and clamp  & \hspace{1cm}Optical Window glass\\ 
& \hspace{1cm}thermal expansion & \hspace{1cm}refractive index change\\ 
& & \\
\multirow{2}{*}{Suspension Struts (SS)} & \hspace{1cm}Optical Bench asymmetric & \hspace{1cm}Optical Bench (OB) \\
& \hspace{1cm}thermal expansion & \hspace{1cm}thermoelastic stress \\
\br
\end{tabular}
\end{center}
\end{table}

\subsection{Thermal Diagnostics Data Analysis}
\label{datanalysis}
A complete model simulating the different consequences of the experiments (temperature responses, dynamics on the TM, and phase shift in the IFO) on one hand, and the telemetry data of the IFO and the temperature sensor readouts, on the other, will be used in the subsequent data analysis, when all the data will be collected and processed so more accurate values can be obtained for the model coefficients. 

\begin{figure}
  \begin{center}
    \includegraphics[width=13cm]{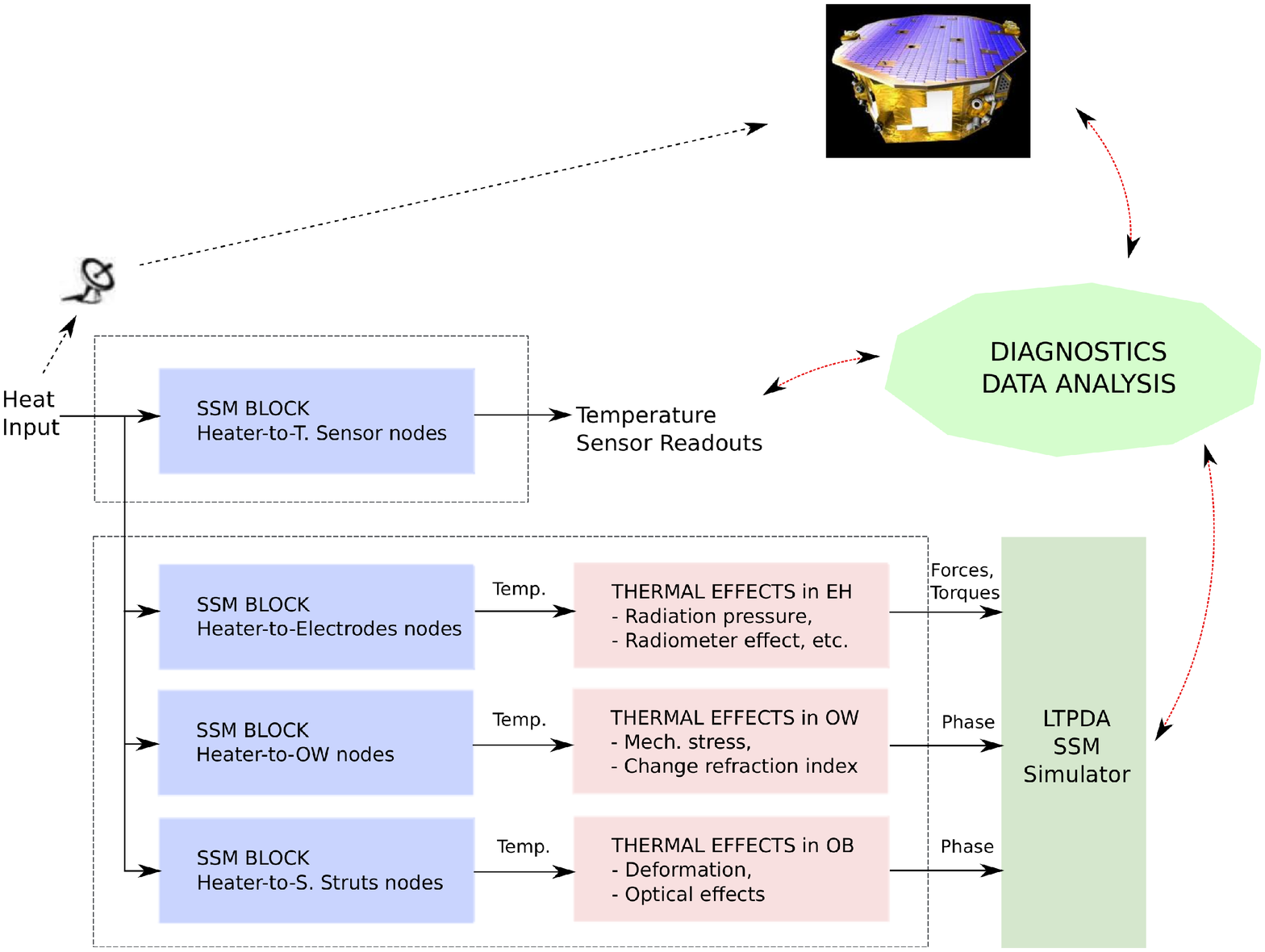}
    \caption{\label{esqtotfig}Scheme of the Thermal Diagnostics model. LPF picture courtesy of ESA.}
  \end{center}
\end{figure}

The global scheme is presented in Figure~\ref{esqtotfig}, where blue blocks refer to the thermal model-obtained temperature, while red ones refer to the blocks representing conversion from temperature distributions to the different perturbations that can disturb the readouts of the interferometer.

The Data Analysis work must handle two different types of simulations: (1) a simulation providing data that will be observed during the mission, as temperature sensor readouts,  and (2) a simulation providing temperatures at non-observable points whose temperatures are inputs to the blocks converting temperature fluctuations into IFO perturbations. Block (1) will be used to validate the thermal model used in (2).

The modelling of the different blocks in Figure~\ref{esqtotfig} varies depending on the feasibility of carrying out experimental on-ground campaigns able to reproduce the disturbing thermal effects of each experiment in Table~\ref{thermexp}. Consequently, experiments for thermal qualification purposes on the Optical Window and on the Optical Bench were planned, the first carried out in AEI Hannover Laboratory (2007,~\cite{owcamp}) and the latter taking place at the time of writing in the frame of the LPF Thermo-Optical Qualification Model (TOQM) campaign in IABG facilities in Ottobrunn, near Munich, being run by EADS Astrium-Germany and EADS Astrium-UK (November 2011, \cite{toqmdoc}). As thermal effects on the free falling TMs can not be directly tested, an accurate model for the EH thermal effects is being developed, the results of which are going to be compared with similar experiments carried out at Trento University with a torsion pendulum~\cite{trentopap}.

\section{Thermal effects in the EH}
\label{thermaleffects}
Temperature differences across the TMs environment can produce differential pressures that turn into net forces and torques on the TMs. Three different thermal effects have been identified~\cite{tesimiquel} as possible mechanisms that will create noticeable dynamic effects on the TMs: 

\subsection{Radiation pressure effect}

The radiation pressure effect is based on the temperature-dependence of the radiation emitted by a surface, following the Stefan-Boltzmann Law. The force applied on a surface {\it j} by the presence of another surface {\it i} at a different temperature is expressed as:
\begin{eqnarray}
    F_{ij}^{RP}=\frac{8}{3}\frac{S_{ij}^{RP}\epsilon_{i} A_j\sigma T_{0}^{3}}{c}\Delta T_{ij}\;\;\;[N] 
    \label{eq.1n1}
\end{eqnarray}
where {\it $\sigma$} is the Stefan-Boltzmann constant, {\it c} is the speed of light, $S_{ij}^{RP}$ is a system geometric factor from surface {\it i} to surface {\it j} which contains information of both the system geometry and potential multi-reflections, $\epsilon_{i}$ is the emissivity of surface {\it i}, $A_j$ is the area of the section where the force is applied, $T_0$ is the absolute temperature and $\Delta T_{ij}$ is the temperature difference between two surfaces, written as $\Delta T_{ij} = T_i-T_j$.

\subsection{Radiometer effect}
The radiometer effect appears in rarefied atmospheres where the particles have a mean free path much longer than the distance between the surfaces of the enclosing volume, i.e in systems with large Knudsen number. The consequent force is represented here as:
\begin{eqnarray}
    F_{ij}^{RM}=\frac{1}{4}\frac{S_{ij}^{RM}A_j p}{T_0}\Delta T_{ij}\;\;\;[N] 
    \label{eq.1n2}
\end{eqnarray}
where {\it p} stands for the remaining gas static pressure in the enclosure and $S_{ij}^{RM}$ is a system geometry coefficient for the radiometer effect, which includes multi-reflections as well. $A_j$, $T_0$ and $\Delta T_{ij}$ have the same meanings as in Equation~\ref{eq.1n1}.

\subsection{Outgassing effect}
The outgassing effect is caused by the detachment of gas molecules from the surface, in a low pressure environment. Its behaviour is strongly related with the surface roughness and the material properties; although its consequences could be similar to the ones from the other thermal effects, it is not yet included in the model because of the uncertainty of some of its parameters.

More details of all these effects can be found in~\cite{tesimiquel} and the values considered for their main parameters are presented in Table~\ref{thermvaltab}, where the radiative emissivities presented there refer to the TM and EH surfaces~\cite{cgsdoc}.

\begin{table}[h]
\caption{\label{thermvaltab}Default values considered for the thermal effects expressions.}
\begin{center}
\begin{tabular}{lll}
\br
Symbol & \hspace{1cm}Parameter & \hspace{1cm}Value \\
\mr
$p$ &\hspace{1cm}Environment static pressure &  \hspace{1cm}$10^{-5}$\,Pa \\
$T_0$ & \hspace{1cm}System temperature  &  \hspace{1cm}293\,K \\
$\epsilon_{TM}$ & \hspace{1cm}TM surface radiative emissivity  &  \hspace{1cm}0.02 \\
$\epsilon_{EH}$ & \hspace{1cm}EH surface radiative emissivity  &  \hspace{1cm}0.05 \\
\br
\end{tabular}
\end{center}
\end{table}

\section{Thermal effects modelling}
\label{modelling}

In this Section, we focus on how the thermal effects inside the Electrode Housing are being modelled. Such a model can be divided into two main blocks, a thermal block that provides temperatures at specific spots when heaters are activated and another block that implements the different thermal effect expressions presented in Section~\ref{thermaleffects} taking into account the geometry of the system and other effects as the potential multi-reflections between surfaces.

\subsection{ESATAN model block}
A thermal model of the LTP in ESATAN software (developed by {\it Carlo Gavazzi Space}) together with a general model of the whole spacecraft is maintained at ESTEC ({\it European Space Research and Technology Centre}). It provides node-to-node transfer function samples, where the input is the heat power applied to a node and the output is the temperature response at another requested node. Temperature responses are calculated as
  \begin{eqnarray}
    T_s(s)=H_{h \rightarrow s}(s) \cdot q_h(s)    
  \end{eqnarray}
where $q_h$ is the heat power applied to heater $h$, $H_{h \rightarrow s}$ the transfer function and $T_s$ the temperature response at sensor $s$, in Laplace Domain.


Different representative nodes in the thermal model were selected, and series of transfer functions relating all the DS heaters to these interesting nodes were obtained and implemented into a single state-space model. 

\subsection{Geometry and multi-reflections considerations}
Since the real system is not composed of infinite parallel surfaces, it is necessary to consider finite size effects when modelling the EH.

\begin{figure}[h!]
\centering
\includegraphics[width=12cm]{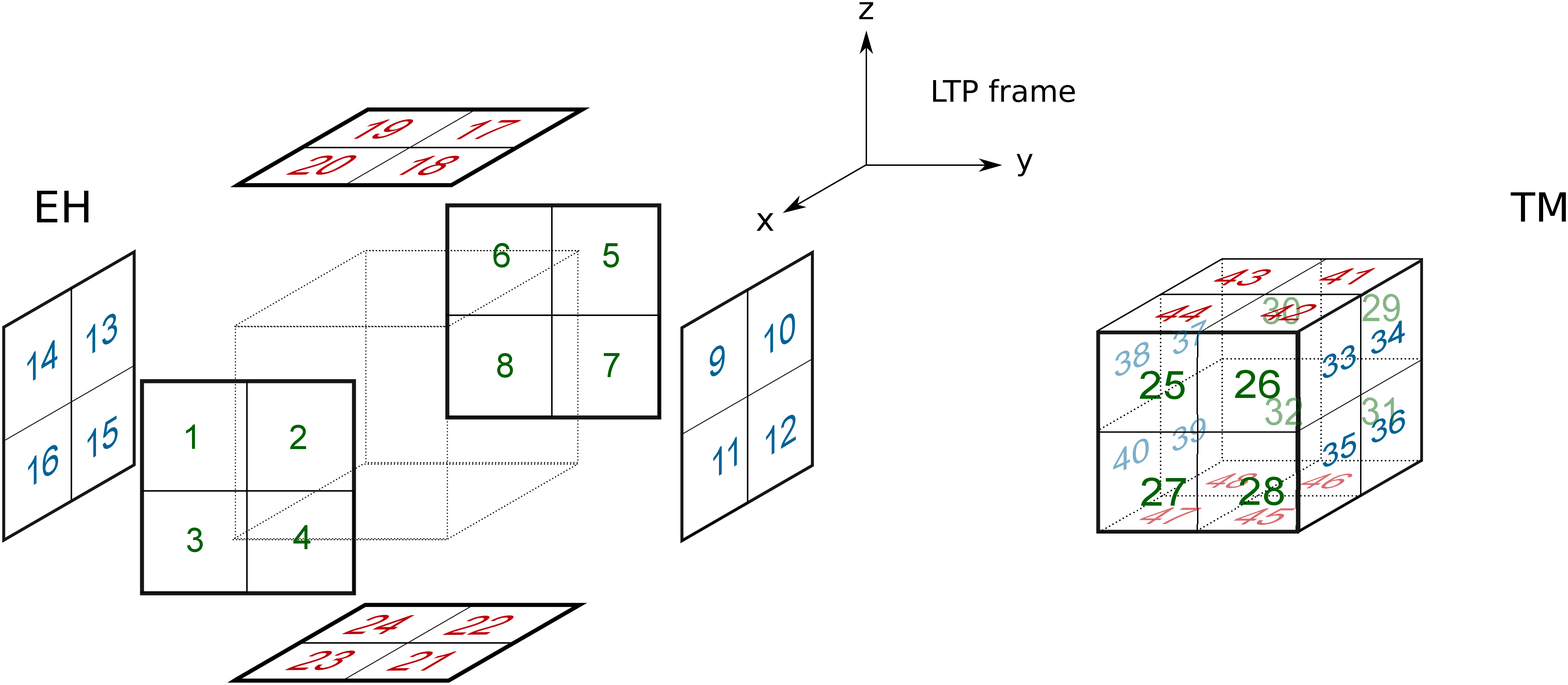}
\caption{Discretisation considered for the temperature distribution in the inner faces of the EH (exploded view) and on the TM with numeration of the facets. Each facet is considered isothermal.}
\label{ehdisc}
\end{figure}

First of all, the surfaces of the system composed of the TM and the inner walls of the EH need to be discretised properly so potential temperature gradients across a surface can be represented. The discretisation adopted is shown in Figure~\ref{ehdisc}, where each TM/EH surface is divided into 4 new facets, thereby enabling the representation of both forces and torques in the three space axis.

On the other hand, in order to model radiation reflection effects in the EH it is necessary to develop a multi-reflection model for our specific system. Additionally, the model must include the possibility of generating not only pressures but also shear forces on the TM surfaces, as they are also important~\cite{trentopap}. 

Such conditions have required the calculation of all facet-to-facet force contributions coming from the different surfaces in a vectorial way (through the computation of the Radiative Exchange Factors~\cite{esatankerneldoc}), so the different resulting components can be disentangled on each surface. The main trade-off of these considerations is the assumption that all reflections are diffuse~\cite{algspec}.

This multi-reflection model is then extrapolated and applied to the Radiometer effect, where equivalent Geometric Exchange Factors have been considered in replacement of the Radiative Exchange Factors, which only apply to the Radiation Pressure Effect. The parameters used in such extrapolations are still under study, but first results are quite satisfactory (see Section~\ref{results}). Table~\ref{refvaltab} presents the current values considered for the different effects and surfaces.

\begin{table}[h]
\caption{\label{refvaltab}Parameters considered in the multi-reflection model. Radiative emissivities have already been presented in Table~\ref{thermvaltab}.}
\begin{center}
\begin{tabular}{lll}
\br
Symbol & \hspace{1cm}Parameter & \hspace{1cm}Value \\
\mr
$\rho_{TM,\,RP}$ & \hspace{1cm}TM radiative reflectivity  &  \hspace{1cm}0.98 \\
$\rho_{EH,\,RP}$ & \hspace{1cm}EH radiative reflectivity  &  \hspace{1cm}0.95 \\
$\epsilon_{TM,\,RM}$ & \hspace{1cm}TM equivalent emissivity for radiometer effect  &  \hspace{1cm}0.02 \\
$\epsilon_{EH,\,RM}$ & \hspace{1cm}EH equivalent emissivity for radiometer effect &  \hspace{1cm}0.05 \\
$\rho_{TM,\,RM}$ & \hspace{1cm}TM equivalent reflectivity for radiometer effect &  \hspace{1cm}0.8 \\
$\rho_{EH,\,RM}$ & \hspace{1cm}EH equivalent reflectivity for radiometer effect &  \hspace{1cm}0.8 \\
\br
\end{tabular}
\end{center}
\end{table}

Finally, as its implementation is linear and it has no dynamics associated, this block can be considered a post-processing part of the temperature state-space model block.

\section{Results}
\label{results}
\begin{figure}[h!]
\begin{center}
\includegraphics[width=10cm]{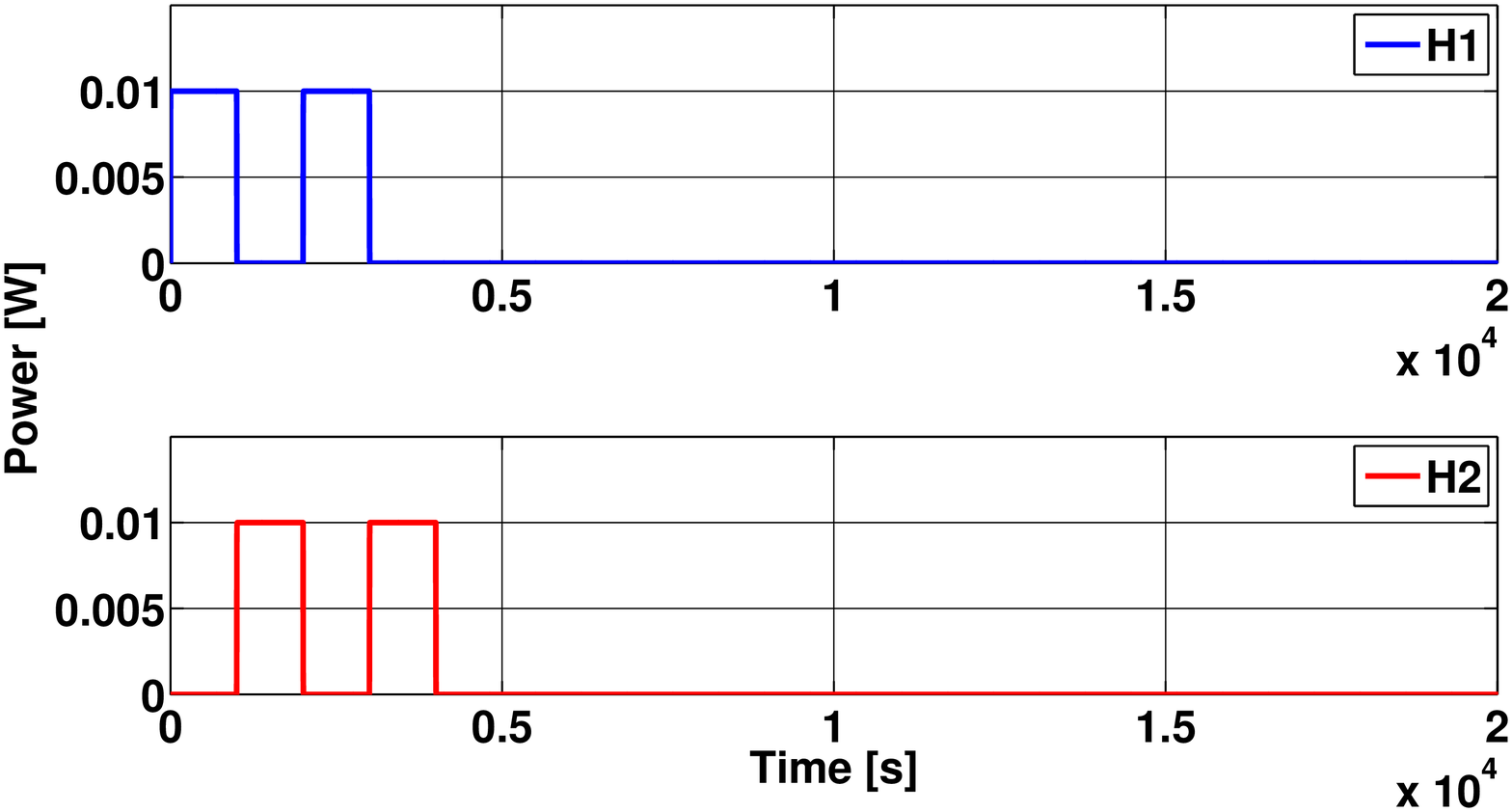}
\caption{Input signals. H1 and H2 are activated alternatively.}
\label{figinput}
\end{center}
\end{figure}

Under the assumption of constant temperature at the TM, we report here first results obtained from the implementation of the model. Specifically, we present forces and torques obtained from the simulation of one planned run from the EMP~\cite{thermalexp}. In this case, the exercise is applied to EH1 and the heaters activated are H1 and H2. The parameters of the input signal are given in Table~\ref{empparams} and Figure~\ref{figinput} shows graphically the input signal shape.

\begin{table}[h]
\caption{\label{empparams}Input signal parameters.}
\begin{center}
\begin{tabular}{ll}
\br
Heaters activated & H1 and H2 alternatively \\
Period & 2000\,s   \\
Duty cycle & 50.00\% \\
Power & 10\,mW (5\,mW to each physical heater) \\
Duration & 4000\,s\\
Total signal duration & 20000\,s \\
\br
\end{tabular}
\end{center}
\end{table}

\begin{figure}[h!]
\centering
\subfigure{
\hspace{-1cm}
\includegraphics[width=9cm]{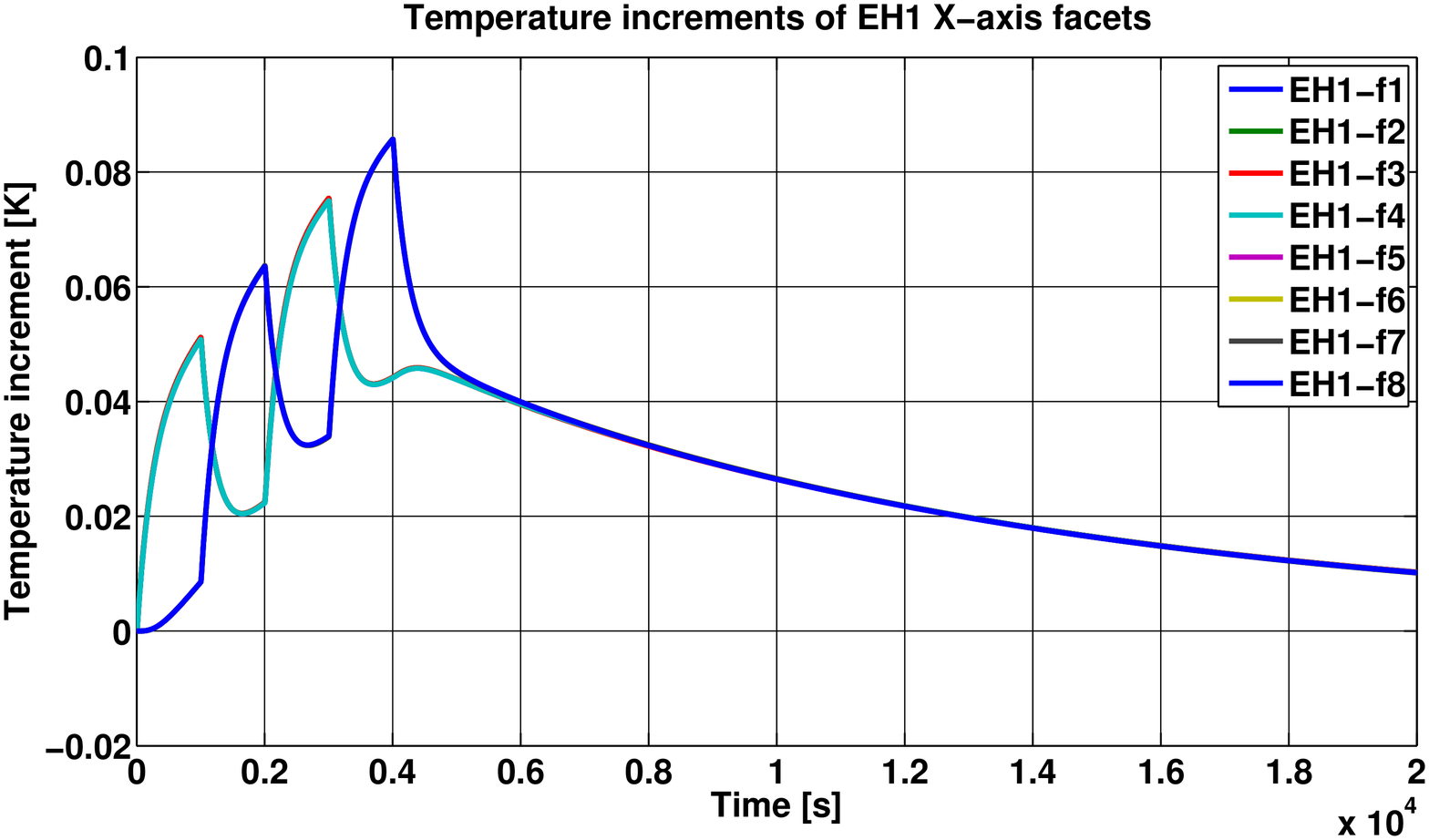}
\hspace{-1cm}
\includegraphics[width=9cm]{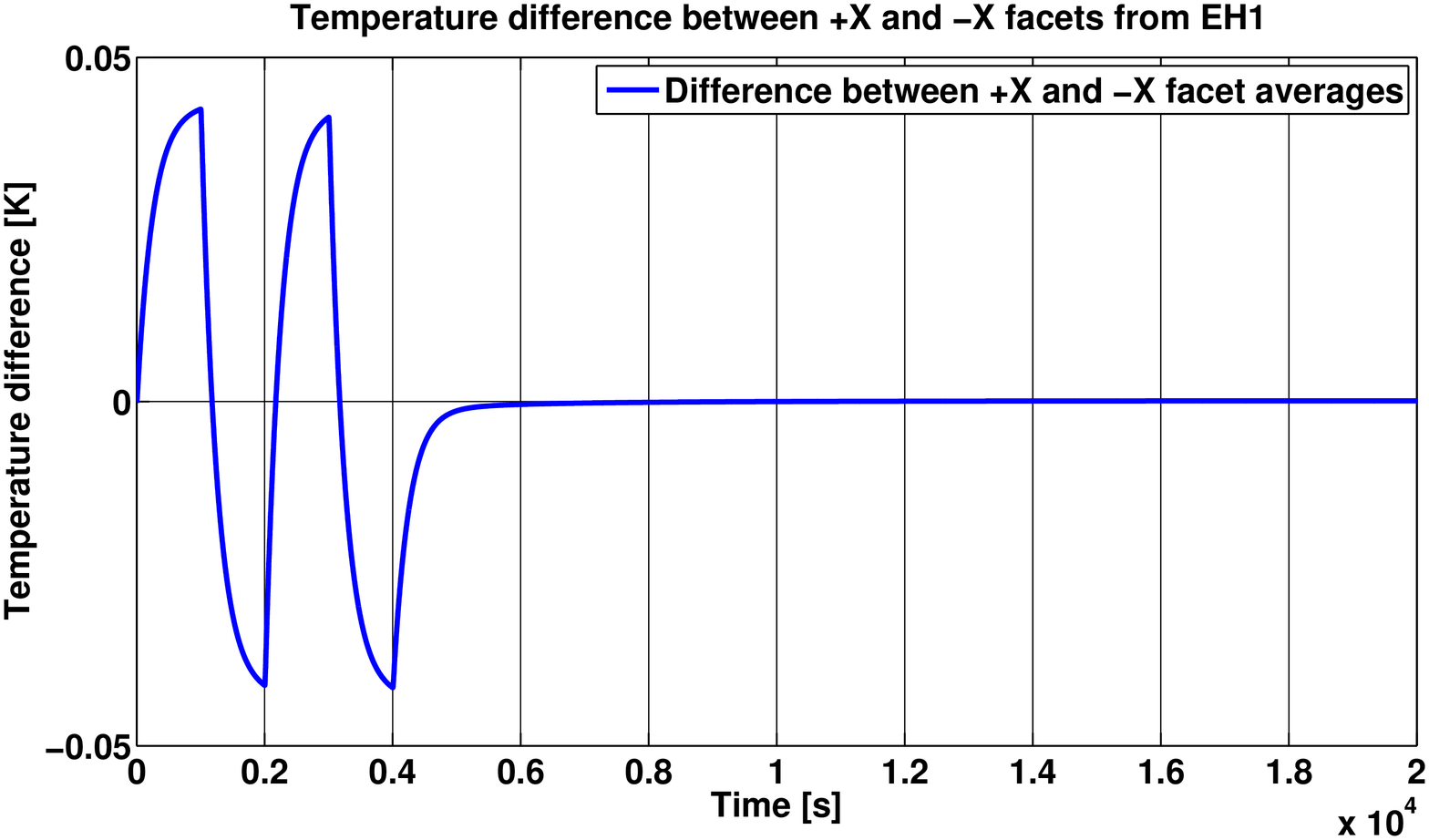}
}
\subfigure{
\hspace{-1cm}
\includegraphics[width=9cm]{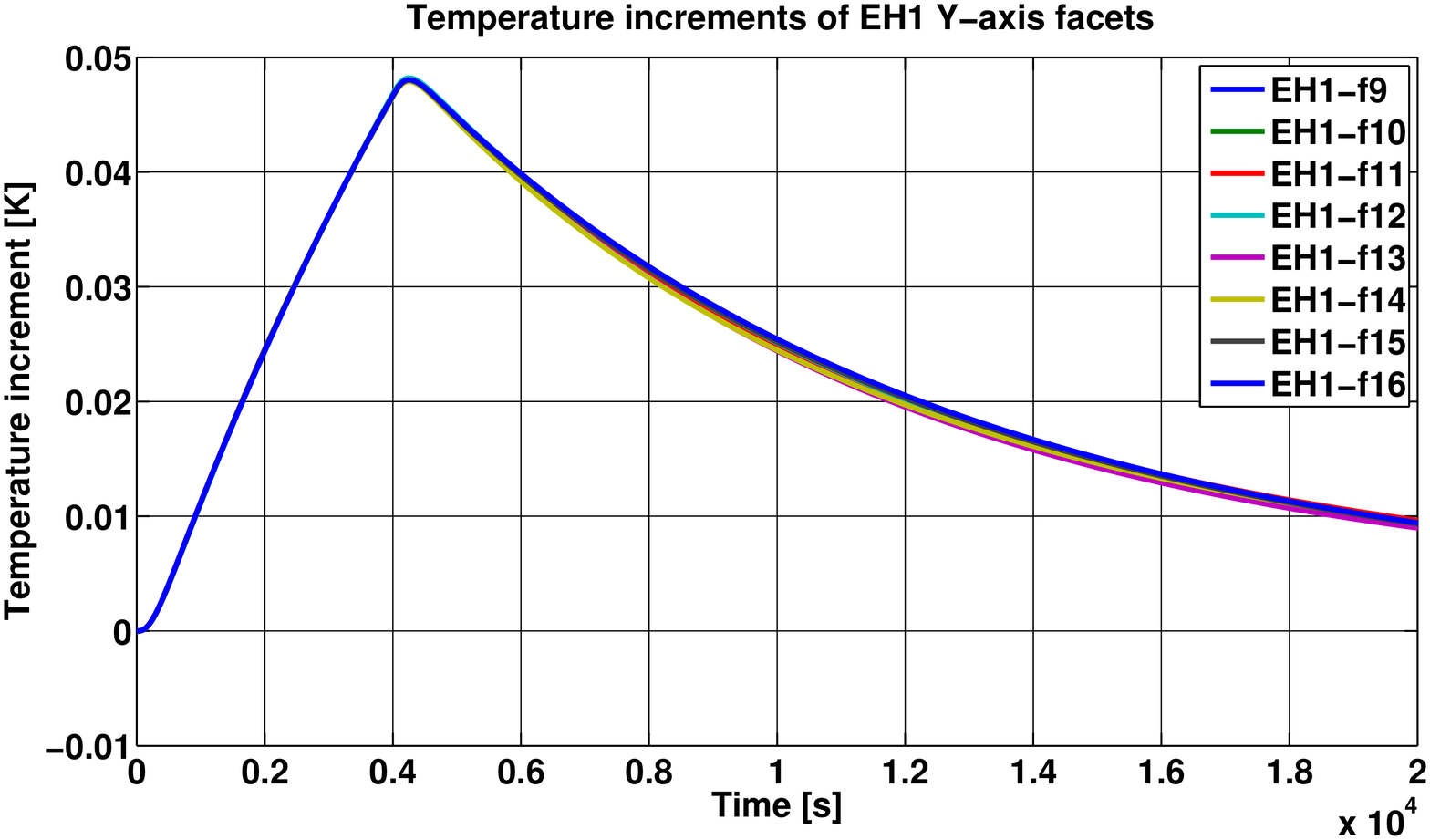}
\hspace{-1cm}
\includegraphics[width=9cm]{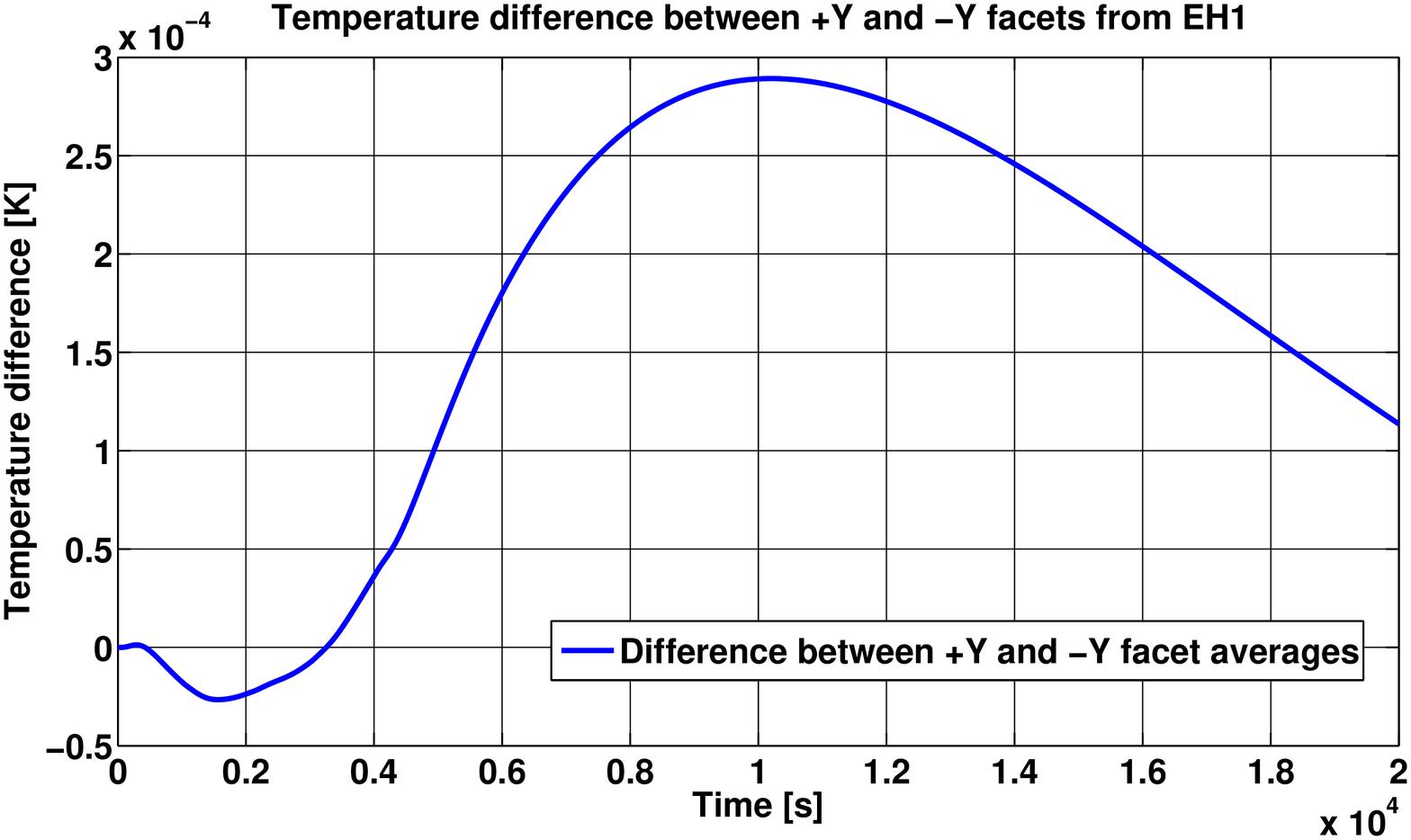}
}
\subfigure{
\hspace{-1cm}
\includegraphics[width=9cm]{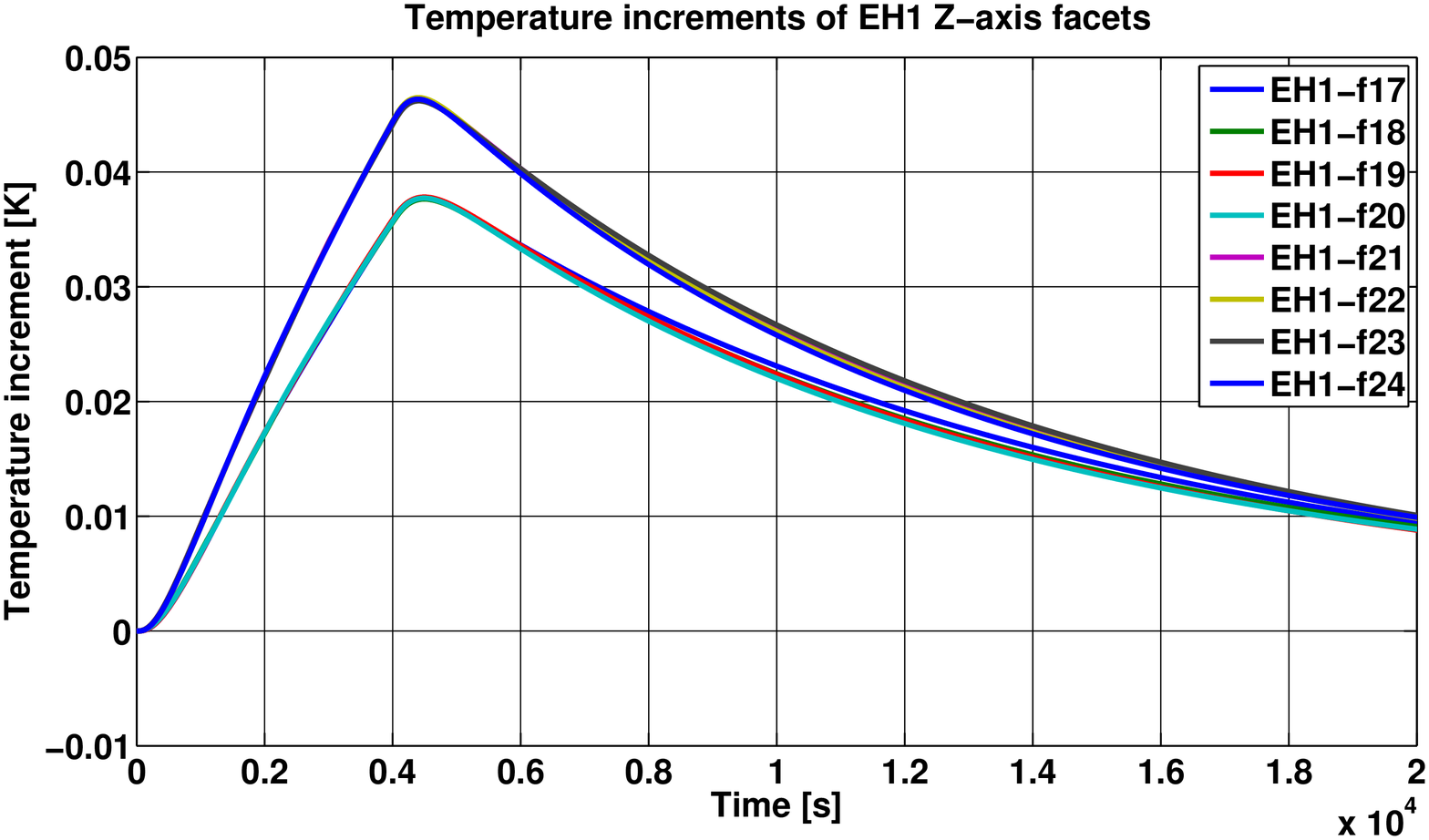}
\hspace{-1cm}
\includegraphics[width=9cm]{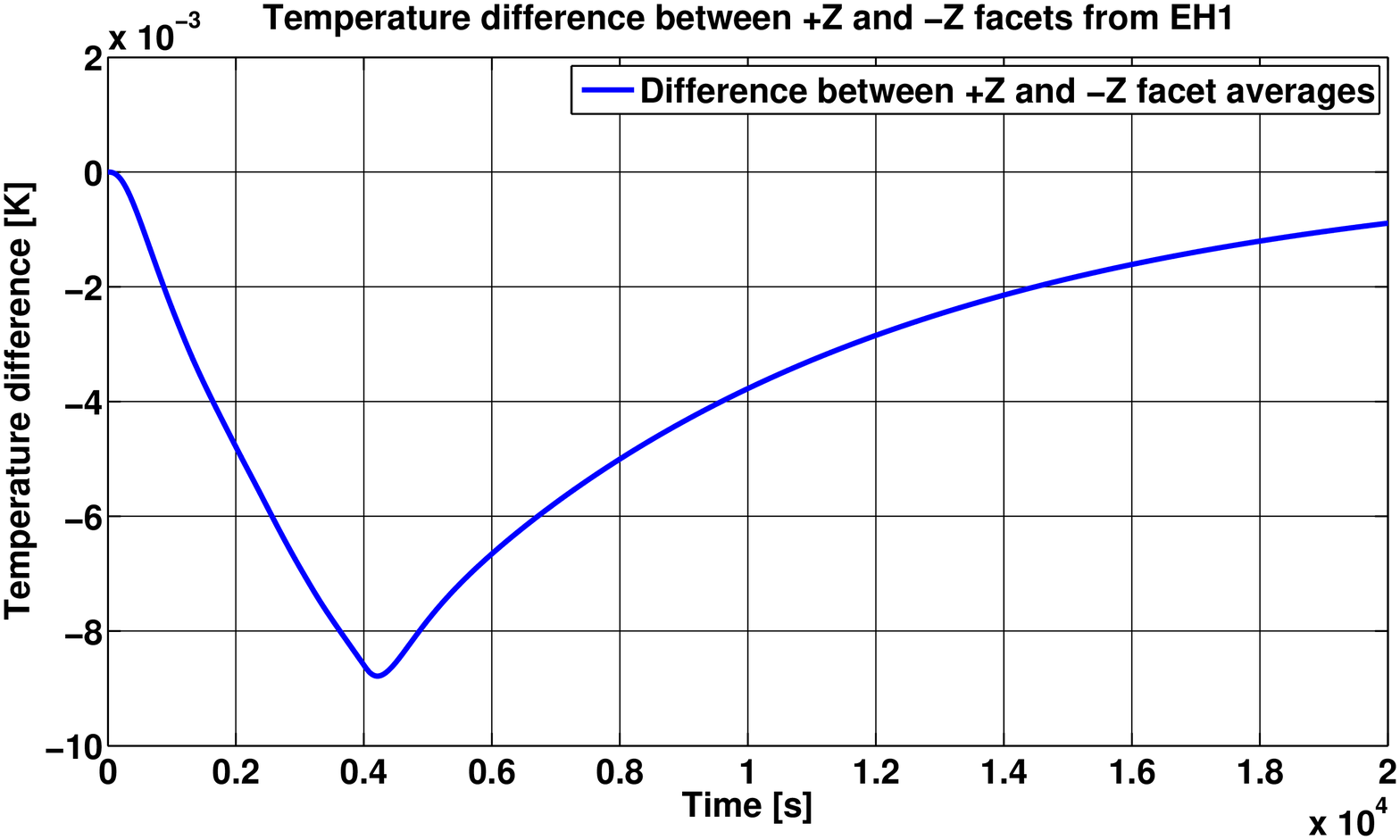}
}
\caption{Plots of temperature increments and differences on EH1 facets. The temperature differences have been obtained by previously averaging the temperature at all the facets from each wall. The facet position follows the notation in Figure~\ref{ehdisc}.}
\label{figtemp}
\end{figure}

Figure~\ref{figtemp} shows the evolution of the EH temperature and the temperature differences, as measured by the sensors in the indicated faces. Temperature sensor readouts in the same face of the EH are not identical, but their differences are negligible in this context.

Note the following temperature evolution features:
\begin{enumerate}
\item When the heater H1 on the +$X$ face is activated, the temperature of the latter starts to rise until the heat input is discontinued, during the second half of the heating signal period. At the same time, the -$X$ face also feels the heating, and its temperature increases, though significantly less.
\item During the second half-period, when H1 is off, the +$X$ face cools down, but does not end up at the original temperature. The activation of H2 in the -$X$ face actually causes the temperature of the +$X$ face to slightly start rising towards the end of the first H2 cycle.
\item As the process continues, there is a consistent drift upwards in the temperature of both faces. When the heaters are completely switched off, the temperatures of the faces match each other -approximately from 5000 seconds-, then they continue in parallel a smooth decline which asymptotically approaches the initial temperature.
\end{enumerate}

In Figure~\ref{figtemp} we see as well plots of the temperature differences between the different faces. As can be observed, a very symmetric pattern shows up in the case of +$X$ and -$X$ faces.

\begin{figure}[h!]
\centering
\subfigure{
\hspace{-1cm}
\includegraphics[width=9cm]{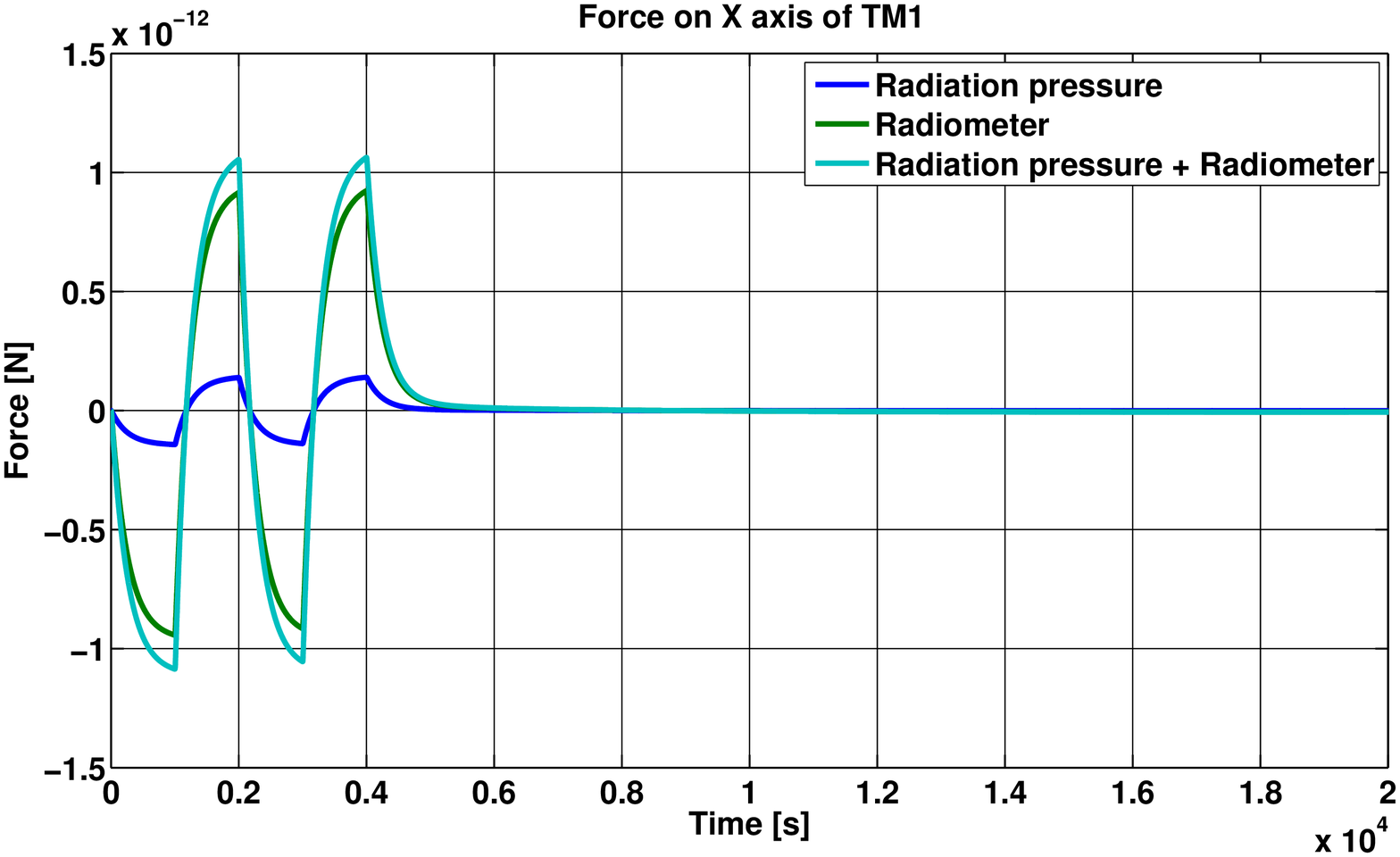}
\hspace{-1cm}
\includegraphics[width=9cm]{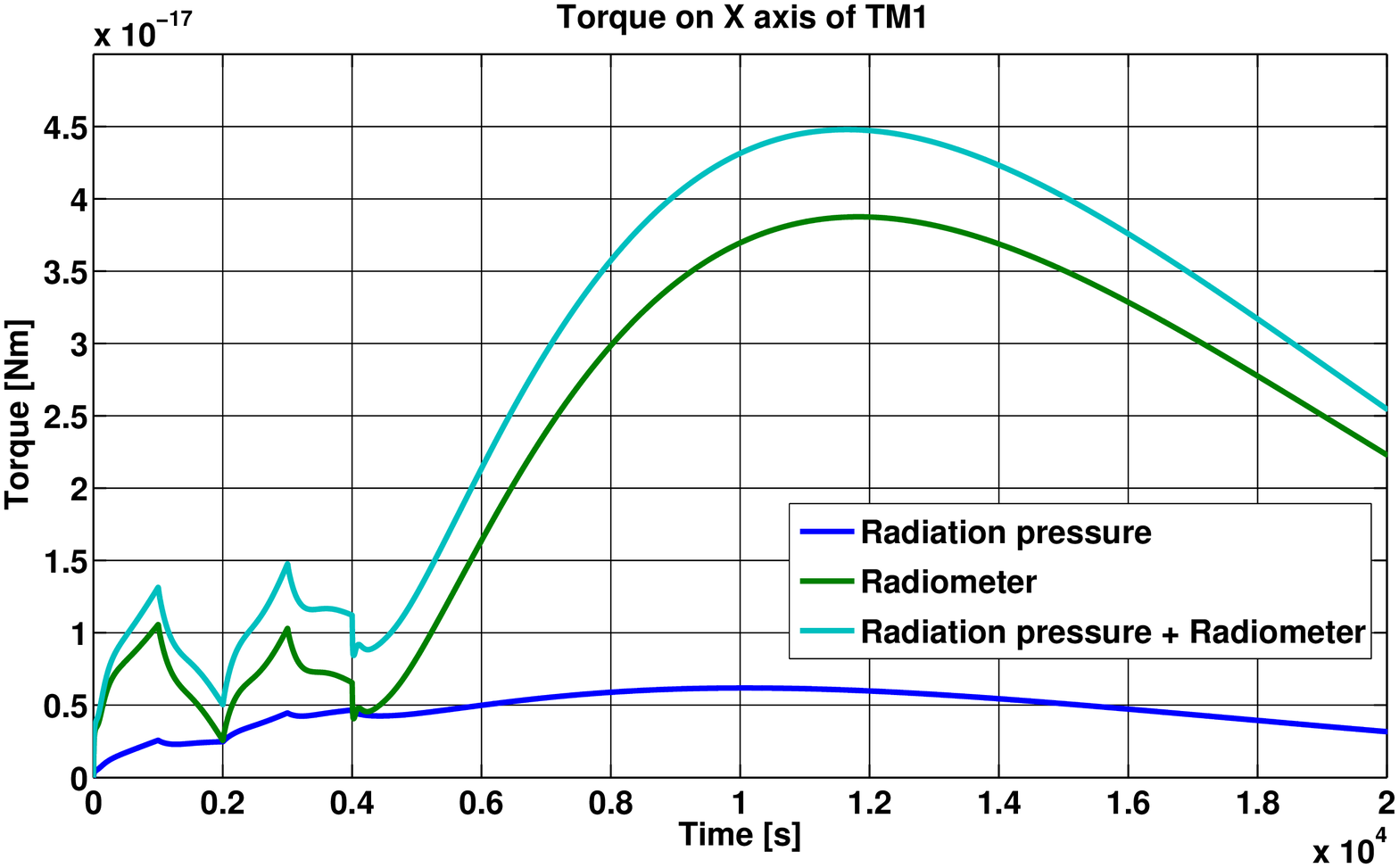}
}
\subfigure{
\hspace{-1cm}
\includegraphics[width=9cm]{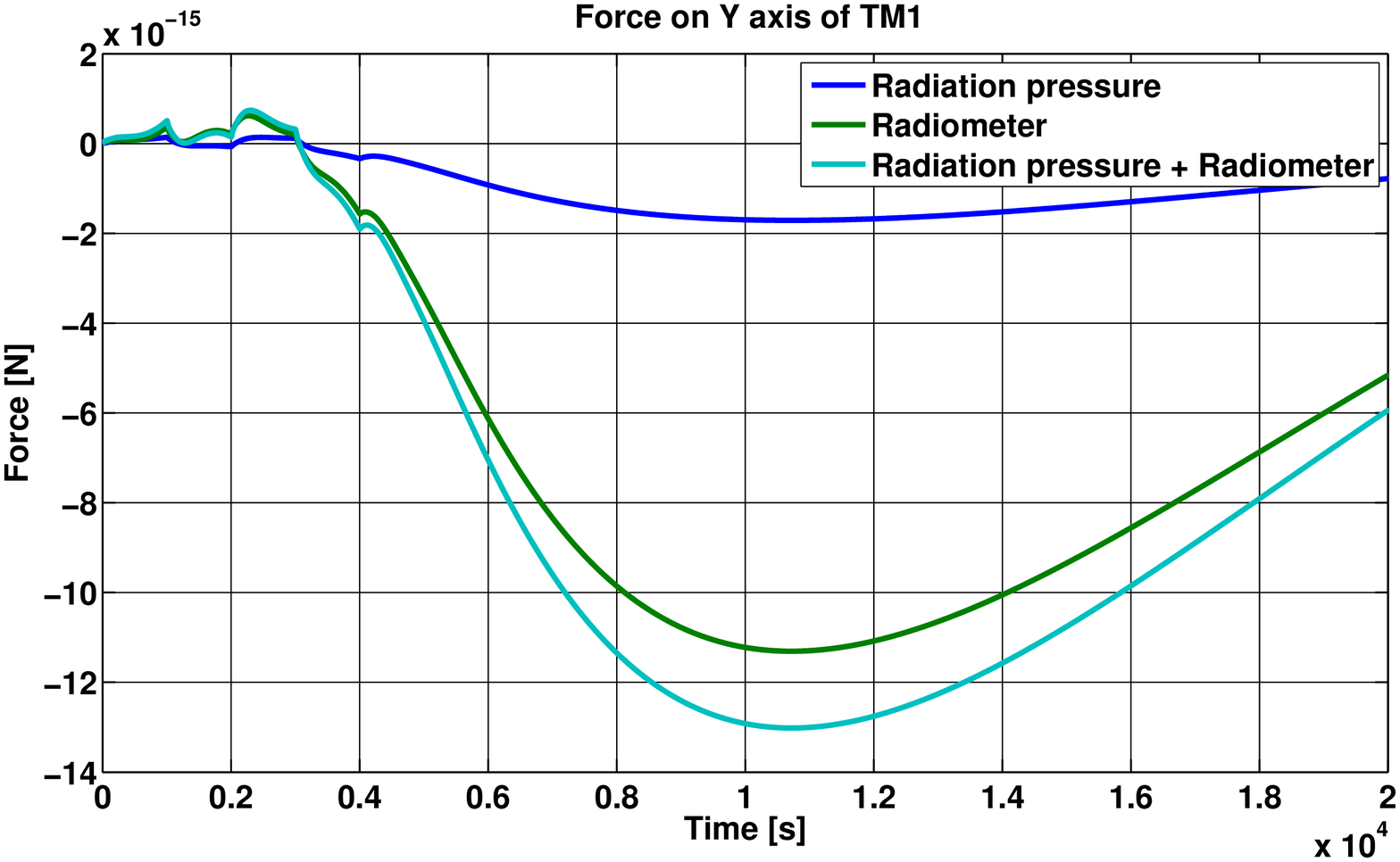}
\hspace{-1cm}
\includegraphics[width=9cm]{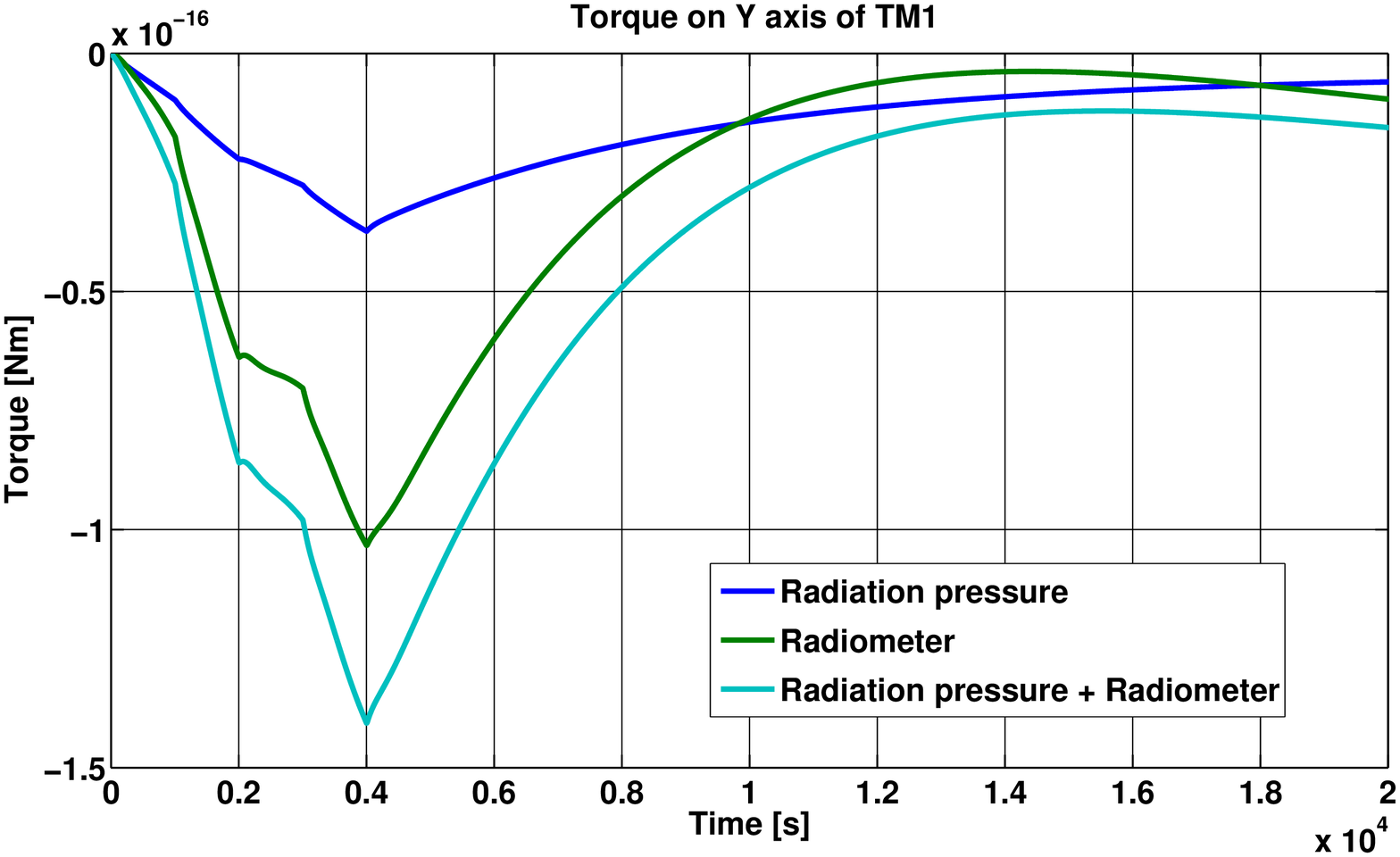}
}
\subfigure{
\hspace{-1cm}
\includegraphics[width=9cm]{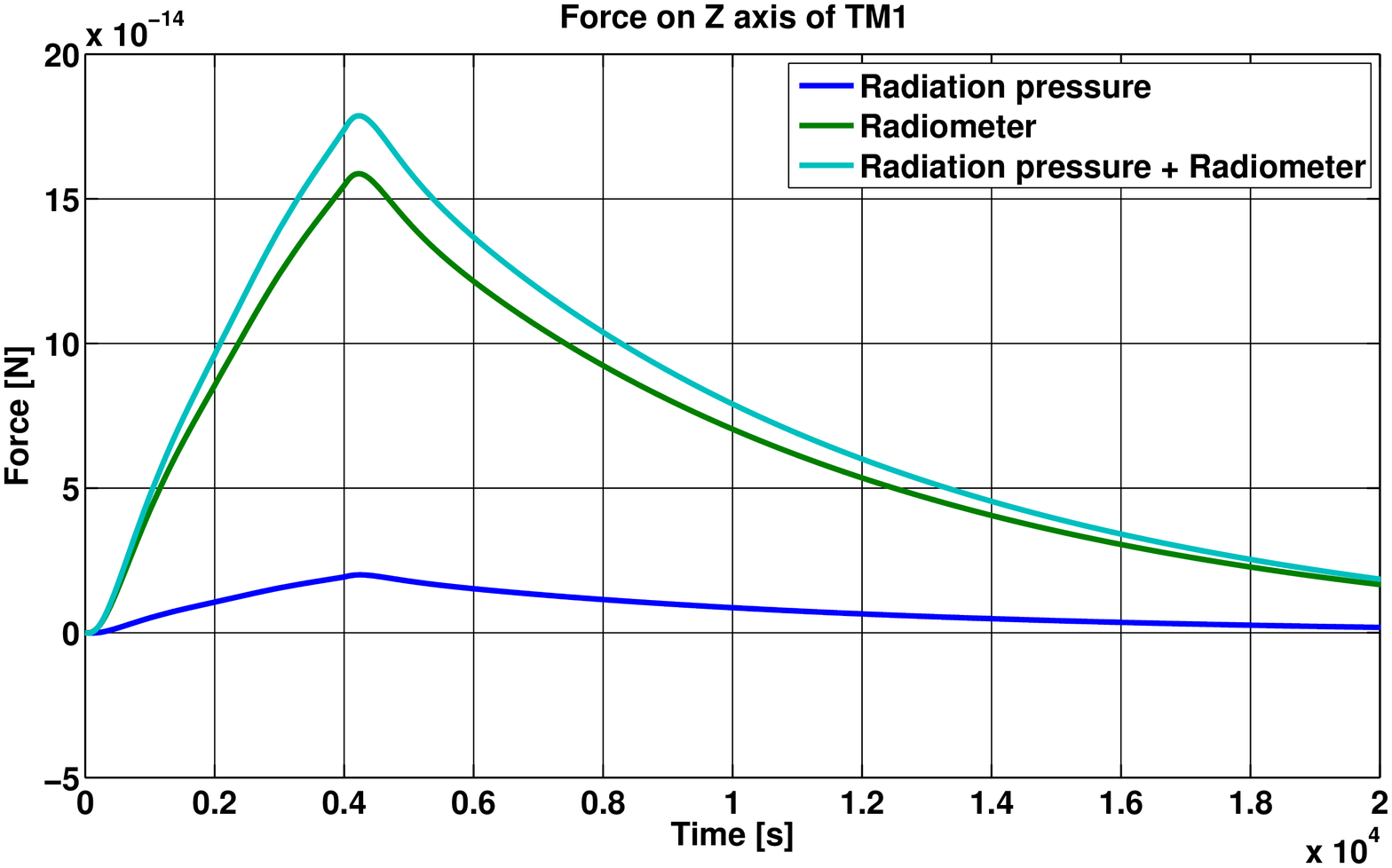}
\hspace{-1cm}
\includegraphics[width=9cm]{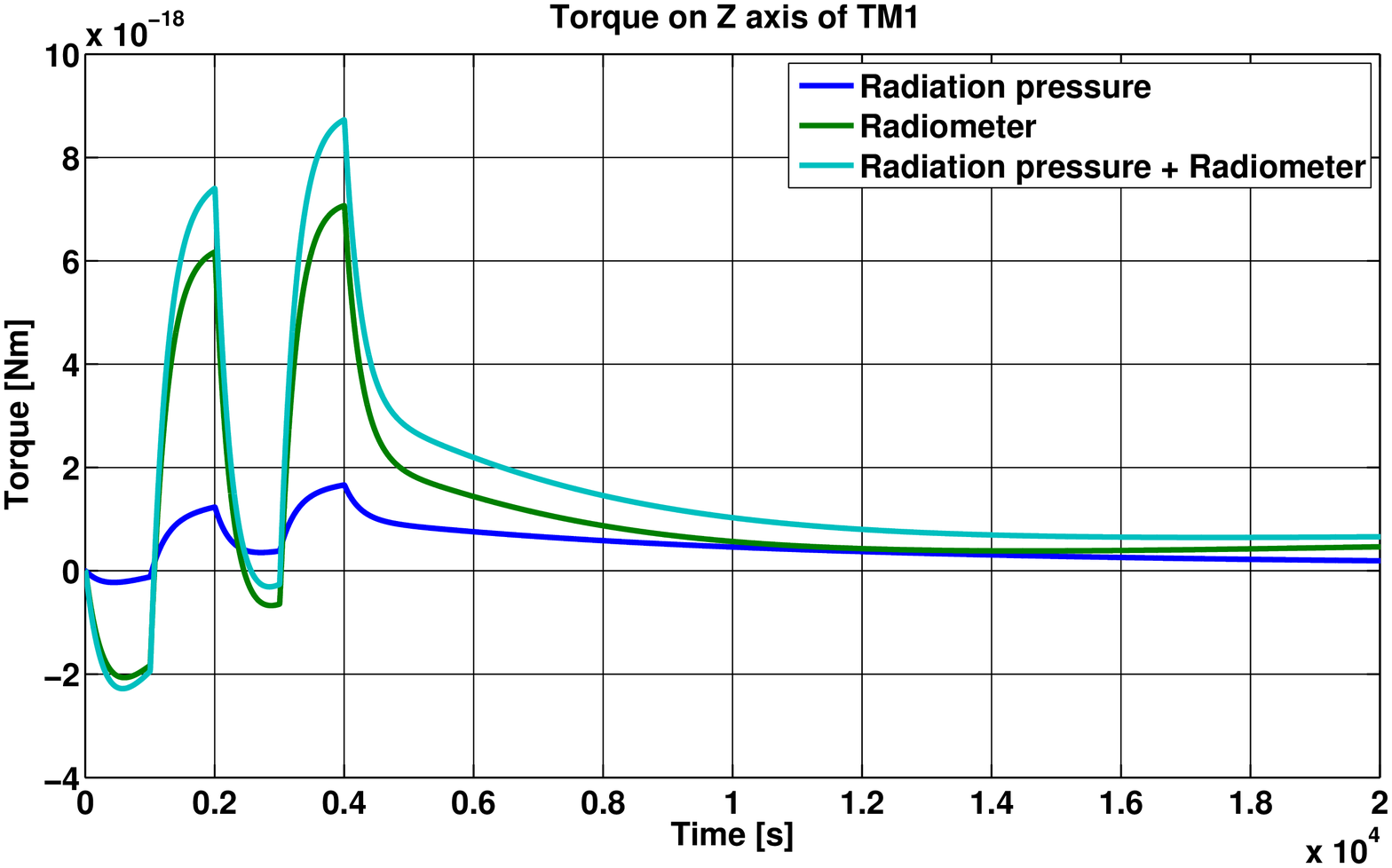}
}
\caption{Plots of forces and torques on TM1, with the disentangled contributions of the different thermal effects under the default parameters of the algorithm.}
\label{figforces}
\end{figure}

Finally, the forces and torques obtained on TM1 considering the contribution of each different effect are presented in Figure~\ref{figforces}. At each plot the contributions of each thermal effect are presented separately, and the parameters applied are the default ones used in the algorithm.

\section{Conclusions}
A model for the thermal Diagnostics Experiments in the Electrode Housing to be carried out in the LPF has been presented. Improved from older versions, this model can now represent multi-reflections for the different thermal effects in a vectorial way, and it includes representations of the radiation pressure and the radiometer effect.

First results show that noticeable thermally-created forces can be obtained in the IFO-sensitive axis -$X$ axis- and that the rest of perturbations, i.e. forces on the $Y$ and $Z$ axis and the different torques, present much smaller effects. On the other hand, all torques observed are negligible as they are far below the DC $1.1\times 10^{-11}\, \rm Nm$ specified in~\cite{reqtor}. Additionally, an interesting temperature gradient is observed on the $Z$ axis, though it is out of the bandwidth of interest.

However, some of the parameters considered for the model are still under on-going study so improved values are expected to be obtained in the future. These parameters refer mostly to the multi-reflections block.

Future work also includes the integration of all the Thermal Diagnostics Experiments into a single Module able to simulate the thermal disturbances in the LTP caused by the different DS heaters, i.e. adding to the current model the OW phase shift and potentially thermoelastic effects on the OB caused by the OW and the Suspension Strut heaters respectively. The implementation of the part involving the EH and the OW in the LTPDA simulator so as to obtain IFO readouts from the thermal experiments is already under way.

\ack We acknowledge support from Project ESP2007-61712 of Plan Nacional del Espacio of the Spanish Ministry of Science and Innovation (MICINN).


\end{document}